\RequirePackage{fix-cm}
\documentclass[twocolumn,epjc3]{svjour3}  
\smartqed  
\RequirePackage{graphicx}
\RequirePackage{comment}
\usepackage{color}
\usepackage{lineno}
\usepackage{url}
\usepackage{amsmath}
\usepackage{breqn}
\usepackage{wasysym}
\usepackage{amsfonts}

\newcommand{\LMO}{Li$_2$MoO$_4$}
\newcommand{\TEO}{TeO$_2$}

\newcommand{\DBD}{$0\nu$DBD}

\begin{document}

\title{Final results of CALDER: Kinetic inductance light detectors to search for rare events}

\author{
L.~Cardani\thanksref{INFNRoma}
\and N.~Casali\thanksref{e1,INFNRoma}
\and I.~Colantoni\thanksref{CNRRoma,INFNRoma}
\and A.~Cruciani\thanksref{INFNRoma}
\and S.~Di~Domizio\thanksref{Genova,INFNGenova}
\and M. Martinez\thanksref{Spagna}
\and V.~Pettinacci\thanksref{INFNRoma}
\and G.~Pettinari\thanksref{IFN}
\and M.~Vignati\thanksref{SAP,INFNRoma}}
\institute{
INFN - Sezione di Roma, Roma I-00185 - Italy\label{INFNRoma}
\and
Consiglio Nazionale delle Ricerche, Istituto di Nanotecnologia  (CNR - NANOTEC), c/o Dipartimento di  Fisica, Sapienza Universit\`{a} di Roma, Roma I-00185 - Italy\label{CNRRoma}
\and
Dipartimento di Fisica, Universit\`{a} di Genova, Genova I-16146 - Italy\label{Genova}
\and
INFN - Sezione di Genova, Genova I-16146 - Italy\label{INFNGenova}
\and
Laboratorio de F\'isica Nuclear y Astropart\'iculas, Universidad de Zaragoza, C/ Pedro Cerbuna 12, 50009 and Fundaci\'on ARAID, Av. de Ranillas 1D, 50018 Zaragoza, Spain\label{Spagna}
\and
Consiglio Nazionale delle Ricerche, Istituto di Fotonica e Nanotecnologie (CNR - IFN), Via Cineto Romano 42, 00156, Roma - Italy\label{IFN}
\and
Dipartimento di Fisica, Sapienza Universit\`{a} di Roma, Roma I-00185 - Italy\label{SAP}
}

\thankstext{e1}{e-mail: nicola.casali@roma1.infn.it}


\date{Received: date / Accepted: date}

\maketitle
\begin{abstract}
The next generation of bolometric experiments searching for rave events, in particular for the  neutrino-less double beta decay, needs fast, high-sensitivity and easy-to-scale cryogenic light detectors. 
The CALDER project (2014-2020) developed a new technology for light detection at cryogenic temperature. In this paper we describe the achievements and the final prototype of this project, consisting of a $5\times5$~cm$^2$, 650~$\mu$m thick silicon substrate coupled to a single kinetic inductance detector made of a three-layer aluminum-titanium-aluminum.
The baseline energy resolution is 34$\pm$1(stat)$\pm$2(syst)~eV RMS and the response time is 120~$\mu$s.
These features, along with the natural multiplexing capability of kinetic inductance detectors, meet the requirements of future large-scale experiments.

\end{abstract}

\section{Introduction}
\label{intro}
More than 60 years after their first detection by Cowan and Reines~\cite{Cowan:1992xc},
the  properties of neutrinos are still being investigated. Among them, the mass scale as well as their nature (Dirac or Majorana particle) are still unknown. 
If observed, a hypothesized nuclear process could unveil both of them: neutrino-less double beta decay ({\DBD})~\cite{Furry,Deppisch:2012nb}. {\DBD} is not allowed by the Standard Model since it violates the conservation of the B-L number~\cite{Vissani:2016}. In this transition a nucleus decays emitting two electrons and no neutrinos, (A,Z)$\rightarrow$(A,Z+2)+2e$^-$, and the signal would consist of two electrons with a total kinetic energy equal to the Q-value of the transition. Current upper limits on the half-life of the decay are of the order of $10^{24}-10^{26}$~yr, depending on the isotope under study~\cite{agostini2020,PhysRevLett.117.082503,PhysRevC.100.025501,PhysRevLett.123.161802,PhysRevD.92.072011,PhysRevLett.123.032501,Adams:2019jhp,Armengaud:2020:cupidmo}. 

Cryogenic calorimeters, hystorically also called ``bolo-meters", are among the leading technologies used to search for {\DBD}. The most sensitive experiment based on this technique, CUORE~\cite{Artusa:2014lgv}, is operating 988 {\TEO} cryogenic calorimeters of $\sim$750~g each with an average energy resolution of $7.8\pm0.5$~keV and a background index of $(1.49\pm0.04)\times10^{-2}$~counts/(keV~kg~yr) in the energy region of interest~\cite{cuore2021}. These performances allowed the CUORE experiment to set a 90\% C.I. Bayesian lower limit of $3.2\times10^{25}$~yr on the $^{130}$Te half-life for {\DBD}.
The CUPID (CUORE Upgrade with Particle Identification) collaboration~\cite{CUPIDInterestGroup:2019inu} proposed a next generation {\DBD} experiment to upgrade the technique of CUORE, with the goal of decreasing the background by two orders of magnitude. CUPID will implement an active particle identification technique in order to identify and reject the dominant background source, i.e. $\alpha$ particles emitted by residual radioactive contamination in the materials facing the detectors~\cite{Alduino:2017qet}. 

In the first design, CUPID was based on TeO$_2$ cryogenic calorimeters, as in CUORE, complemented by cryogenic light detectors to detect the tiny amount of Cherenkov light ($\sim$100~eV) emitted only by $\beta/\gamma$ interactions~\cite{Casali:2016luq}, thus enabling the background rejection. Nevertheless, the results obtained by CUPID-0~\cite{PhysRevLett.123.032501,Azzolini:2018tum,PhysRevLett.123.262501,Azzolini_excited_states,Azzolini_Zn_decay,PhysRevD.100.092002} and CUPID-Mo~\cite{Armengaud:2020:cupidmo,Armengaud:2019loe} experiments demonstrated that the $\alpha$ particles rejection capability can be more easily achieved by using scintillating crystals like ZnSe (\DBD\ of $^{82}$Se) and {\LMO} (\DBD\ of $^{100}$Mo) in place of the {\TEO} ones because of their higher light signal compared with the Cherenkov yield. Furthermore $^{82}$Se and $^{100}$Mo benefit of a lower intrinsic background with respect to $^{130}$Te because their {\DBD} Q-value is above 2615~keV, where the $\beta/\gamma$ background contribution coming from natural radioactivity is greatly reduced. The CUPID group therefore decided to replace {\TEO} crystals in favour of ZnSe or {\LMO}. Finally, because of the better radio-purity and energy resolution with respect to ZnSe, {\LMO} was chosen as crystal for CUPID.

It is clear from the previous considerations that the light detection technology will play a key role in the future of CUPID. To reach the background goal of $10^{-4}$ counts/ (keV~kg~yr), light detectors with small baseline fluctuations ($<100$~eV RMS), good time resolution (better than $1$~ms) and high reproducibility are needed. The light detectors of the first phase of the project  will be based on the well established technology of germanium wafers equipped with Neutron Transmutation Doped Germanium (NTD-Ge) thermistors~\cite{Beeman:2013zva}.
These sensors offer a typical baseline resolution of 50--60~eV RMS with rise time of 1--5~ms (see, e.g., Refs.\cite{Armengaud:2019loe,Artusa_2016,Ressa_2021} and references therein). A detector of 16~cm$^2$ in an optimized configuration reached an ultimate resolution of 20\,eV RMS with rise time of 0.8~ms~\cite{Barucci:2019ghi}.

Nevertheless, the physics reach of CUPID (or other bolometric projects such as AMoRE~\cite{Alenkov:2019jis}), could be largely increased with faster light detectors. Rise times of the order of hundreds of $\mu$s or below would allow to reject the ultimate envisioned background, consisting of pile-up events from the naturally occurring double-$\beta$ decay with the emission of 2 neutrinos~\cite{CUPIDInterestGroup:2020rna,Chernyak2017}. Furthermore, detectors offering multiplexing capabilities would allow to increase the number of devices without increasing the thermal load on the cryogenic facility, and ease of fabrication and low-cost readout are highly desirable features for large scale experiments. Apart from the limited time resolution, NTD sensors, being high-impedance devices, cannot be easily  multiplexed at cryogenic temperatures.

Light detectors based on transition edge sensors (TES) have also been proposed as an alternative to NTDs for rare events searches~\cite{Schaffner:2014caa} and demonstrated an extraordinary energy resolution down to 3.9~eV RMS with a distributed network of sensors~\cite{Fink}. Nevertheless, these devices cannot be easily scaled to thousands of channels because of their complex readout (based on SQUIDs) and limited reproducibility.

In this work we summarize the R\&D and present the final results of the CALDER project, which developed an alternative technology to NTDs and TESs for the light detection based on Kinetic Inductance Detectors.


\section{Kinetic Inductance Detectors with phonon mediation}
\label{sec:KID}
\begin{figure}[t]
\begin{centering}
\includegraphics[width=\linewidth]{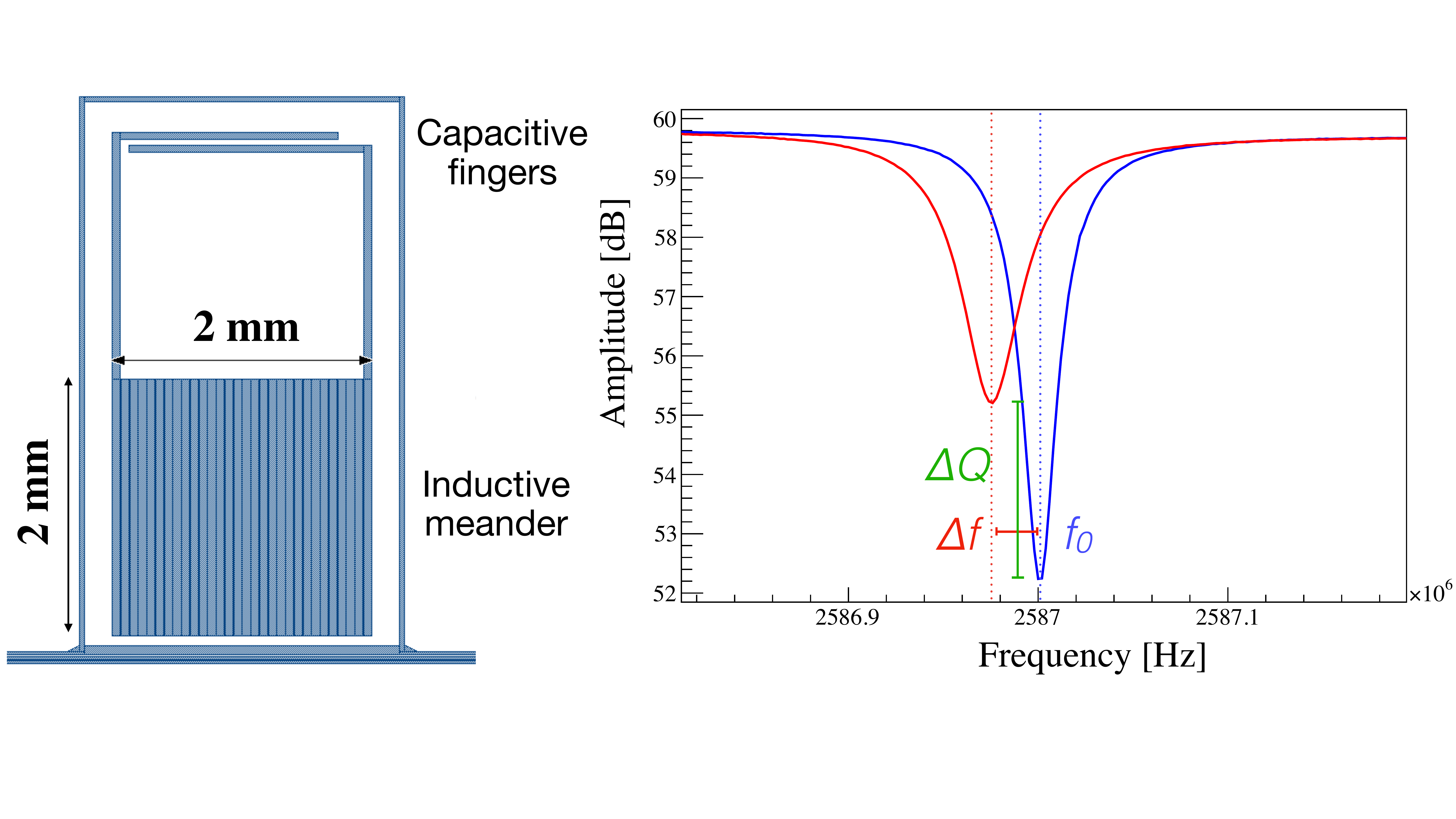}
\caption{Left: KID geometry developed in CALDER. The meandered inductor ($L$) (30 strips of 62.5~$\mu$m$\times$2~mm separated by a gap of 5~$\mu$m) has an active area of around 4~mm$^2$, is closed on a 2-finger capacitor ($C$) on one side, and coupled to a coplanar wave guide (CPW) on the other side. A ground line surrounds the LC circuit for electro-magnetic isolation;
Right: Frequency scan of a KID. When energy is absorbed in the inductor, the resonant frequency $f_0$ and quality factor $Q$ of the resonator are reduced. When the KID is biased on-resonance at fixed readout frequency, the signal is detected as phase and amplitude modulations of the wave transmitted past the device.}
\label{fig:KID}
\end{centering}
\end{figure}

In 2003 a new technology was proposed for the detection of electromagnetic radiation at cryogenic temperatures: the Kinetic Inductance Detectors (KIDs)~\cite{bay}. A KID consist of a resonant LC circuit made of a superconducting metal operated well below its critical temperature $T_c$ (see the layout of the KID of CALDER in Fig.~\ref{fig:KID} left). The inductance $L$ is  the sum of the magnetic inductance ($L_g$), which depends on the geometry of the inductor, and of the so called kinetic inductance ($L_k$), which depends on the density of Cooper pairs in the superconductor.

For common superconductors such as aluminum, the binding energy of the Cooper pairs is of the order of hundreds of $\mu$eV, so that even a tiny amount of absorbed energy can break a high number of Cooper pairs. The pair-breaking causes an increase of $L_k$ and of the surface resistance of the superconductor, $R$, resulting in a measurable change of the frequency ($f_0 = 1/2\pi\sqrt{LC}$) and the quality factor ($1/Q=R\sqrt{C/L}$) of the resonator (Fig.~\ref{fig:KID} right). 

What makes KIDs interesting with respect to other cryogenic sensors is their natural multiplexing in the frequency domain. Indeed, hundreds of KIDs can be coupled to the same feedline, and can be simultaneously read by making them resonate at slightly different frequencies, which can be obtained by adjusting the layout of the capacitor and/or of the inductor of the circuit~\cite{Mazin:2010pz}. This allowed to successfully operate very large arrays of KIDs, up to 1000~\cite{Adam:2017gba}. Despite these outstanding performances, their application in other  fields  could be limited by their small active surface (few mm$^2$). However, Swenson et al.~\cite{Swenson:2010} and Moore et al.~\cite{Moore:2012} showed that KIDs are also good phonon sensors: X-rays or, more in general, ionising radiation interacting in a substrate generate phonons that, through their scattering in the lattice, can eventually reach the KIDs, break the Cooper pairs and generate a signal. With the mediation of the phonons in the substrate it is therefore possible to increase the sensitive area beyond the KID dimensions, at the cost of a lower sensitivity due to inefficiencies in phonon transmission and absorption in the KID. 

\section{The 4 and 25~cm$^2$ light detectors of CALDER}
\begin{figure}[t]
\begin{centering}
\includegraphics[width=\linewidth]{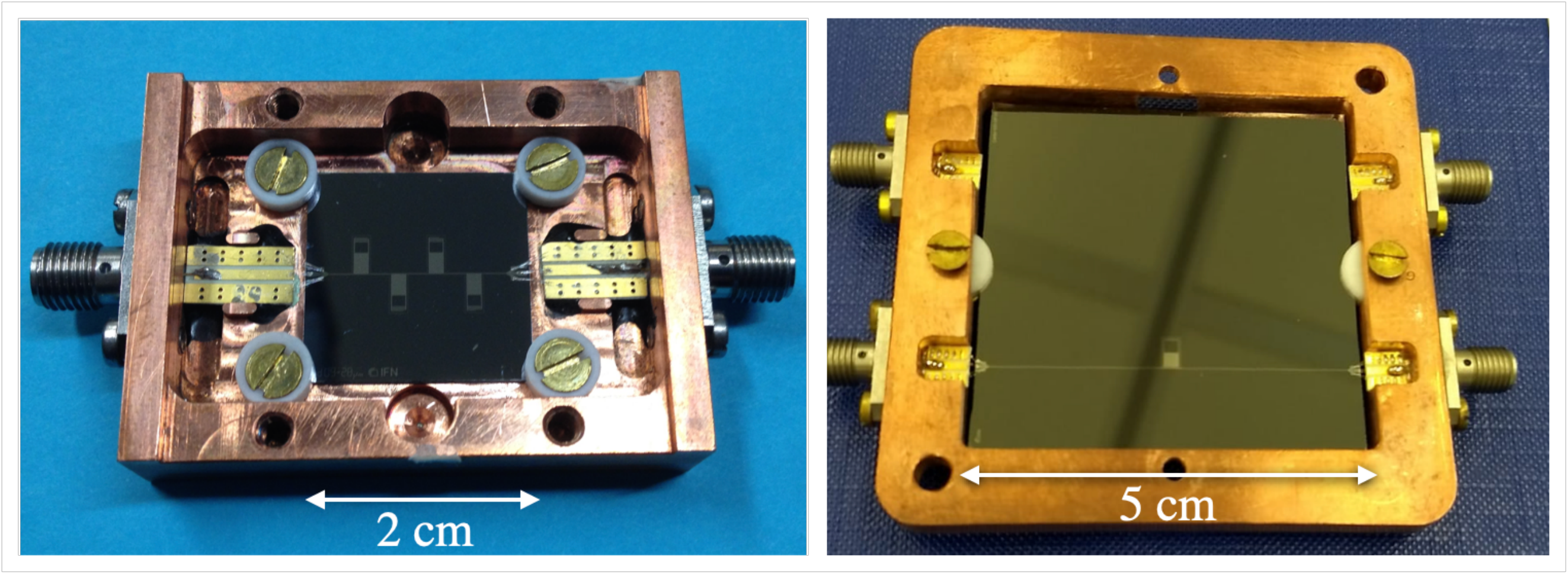}
\caption{Left: first prototype of light detector with 4 KIDs coupled to the same feedline and deposited on a 4~cm$^2$, 300~$\mu$m thick, silicon substrate (Picture adapted from Ref.~\cite{Cardani:2015tqa}). Right: final prototype with 1 KID deposited on a 25~cm$^2$, 650~$\mu$m thick, substrate. }
\label{fig:chips}
\end{centering}
\end{figure}

The CALDER (Cryogenic wide-Area Light Detectors with Excellent Resolution) project~\cite{Battistelli:2015vha} developed large area cryogenic light detectors based on KIDs exploiting the phonon-mediation. The first prototypes~\cite{Colantoni177:2016,Colantoni131:2016} were made depositing 4 multiplexed aluminum resonators on a $2\times2$~cm$^2$, 300~$\mu$m thick silicon substrate (Fig.~\ref{fig:chips} left) and reached an energy resolution of 150 eV RMS~\cite{Cardani:2015tqa}. Then we improved the layout of the LC circuit in order to increase its area for a better phonon collection (from 2.4  to 4 mm$^2$) and in order to increase the quality factor of the resonator (from 15000 to 150000), which in turn needed also an improvement of the quality and accuracy of the metal deposition~\cite{Colantoni726:2018}. Good results were obtained, 
both in terms of energy resolution and rise time (80~eV and 10~$\mu$s respectively) and also in wide range of operating temperatures ($10-200$~mK)~\cite{Bellini:2016lgg}. A similar prototype was successfully used as cryogenic light detector coupled to a $2\times2\times2$~cm$^3$ {\LMO}, demonstrating very promising performances~\cite{Casali:2019baq}.

In order to further improve the energy resolution, we moved the R\&D to KIDs made of superconductors more sensitive than aluminum. Indeed, the sensitivity depends strongly on $L_k$ that in turn depends on the material~\cite{zmu}. At the same time the superconducting material needs to feature a good acoustic match to the silicon, to ensure an efficient phonon transmission and absorption in the KID. 
We started testing several resonators made of sub-stoichiometric titanium nitride without obtaining satisfactory results, mainly because of the poor reproducibility of the film, a difficulty encountered also by other groups~\cite{Leduc:2010,Szypryt:2017}. 
We then moved to granular aluminum that, because of its tunable energy gap and large fraction of kinetic inductance, is a promising material in applications for quantum circuits~\cite{Gruenhaupt2019}. Nevertheless, we were not able to reach a baseline resolution better  than $\sim$150~eV, likely because of the acustic mismatch between granular aluminum and silicon, that prevents an efficient collection of phonons.
On the contrary, multi-layers of aluminum and titanium  showed from the beginning very promising results, thanks to the higher kinetic inductance of titanium.
We tested several bi-layers consisting of titanium and aluminum with variable thickness. The best baseline resolution, 50 eV RMS, was obtained with 10 nm of titanium and 25 nm of aluminum. A further improvement was achieved moving from a bi-layer to a tri-layer of aluminum-titanium-aluminum, in order to enhance the acoustic match between KID and silicon.
The best performance was obtained with a three-layer (14~nm Al, 33~nm Ti, 30~nm Al) that, thanks to a 8 times higher kinetic inductance than aluminum, allowed a very competitive baseline energy resolution of 26~eV RMS~\cite{cruciani2018}. 

The last step of the project consisting in the scale-up of the Si substrate to the final size of $5\times5$~cm$^2$. Such dimensions were chosen at the beginning of the CALDER R\&D activity to match the size of the faces of the TeO$_2$ crystals of CUORE, in order to maximize the light collection efficiency~\cite{Casali:2016luq}. 
Scaling the substrate surface from 2$\times$2 to 5$\times$5 cm$^2$, and at the same time preserving the baseline resolution, turned out to be the most complex technological challenge.
Even if we used the same KID design tested in previous detectors, we had to build and test tens of different prototypes before obtaining a comparable quality factor of the resonator. The reason behind this problem turned out to be the CPW feed line. Indeed, in order to decrease the number of phonons absorbed by non sensitive Al regions its width was only 85~$\mu$m. At the same time, in order to collect as many phonons as possible, we increased the number of KIDs from 1 to 4, with the feedline running through the whole substrate and being 6 times longer than in the $2\times2$~cm$^2$ detector. This caused impedance mismatches that reduced by more than an order of magnitude the quality factor. This effect was not visible in EM simulations and therefore we proceeded by trials and errors. In the end the simplest solution turned out also to be the most reliable and high-performing: we minimized the feedline length by making it straight from one side of the wafer to the opposite one (5~cm), and coupled to it a single KID. Figure~\ref{fig:chips} (right) shows the final detector held in a copper frame by PTFE washers consisting in a $5\times5$~cm$^2$, 650~$\mu$m thick Si wafer monitored by an AlTiAl (14~nm, 33~nm, 30~nm) three-layer resonator with the same geometry showed in Fig~\ref{fig:KID} left. 

\section{Measurement set-up}
\label{sec:detector}
\begin{figure}[t]
\begin{centering}
\includegraphics[width=\linewidth]{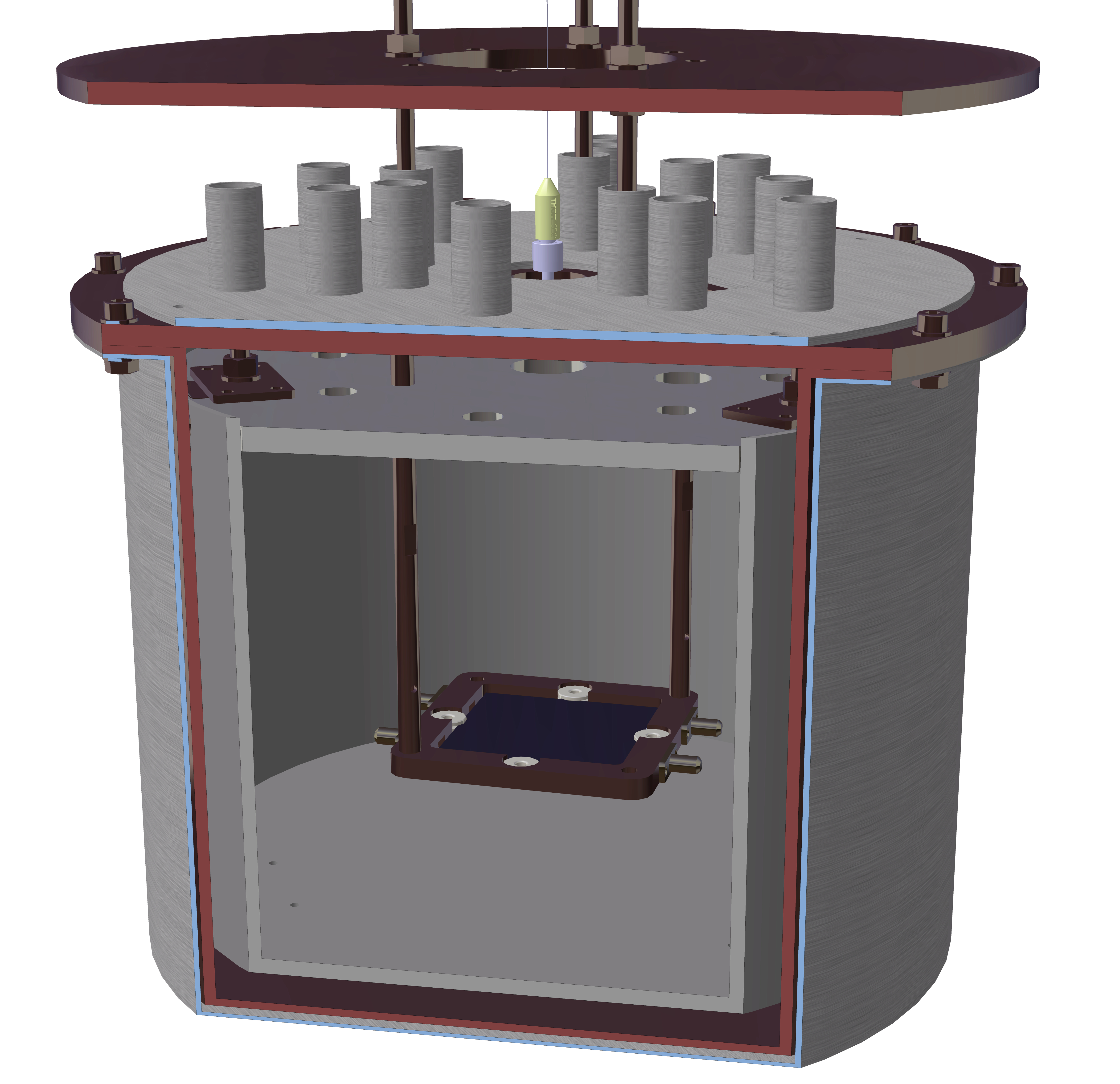}
\caption{Rendering of the shielding pot anchored to the 20~mK plate of the cryostat  at the top. The pot consists of three concentric layers acting as magnetic and radiation shields. The outer one is made of Cryophy\textregistered\ for magnetic shielding, the middle one of copper for thermalization, and the inner one of aluminum for further magnetic and radiation shielding. The mechanical design of the parts constituting the assembly has been carried out considering the effect of the differential thermal contractions related to the different materials constituting the system (copper, aluminum and Cryophy\textregistered). In this way, at cryogenic temperatures, the system has maintained minimum gaps without introducing unwanted mechanical stress on the components. The ``chimneys'' are feedthroughs for cables made of  Cryophy\textregistered as well. The optical fiber, shown in yellow, illuminates the detector from the top and is connected on the other end to a room temperature LED (not shown in the figure). The detector frame is anchored to the copper shield with two copper bars for thermalization.}
\label{fig:pentola}
\end{centering}
\end{figure}
The device  was anchored to the coldest point of a dry $^3$He/$^4$He dilution refrigerator\footnote{Oxford Instruments, Dry Dilution Refrigerator Triton 200.} and cooled down to 20~mK. The outside vessel of the refrigerator was surrounded by a $\mu$-metal cylinder shield in order to reduced the magnetic field at the sample location. Finally, to prevent that both the residual magnetic field inside the cryostat and the thermal radiation from the 600~mK stage spoiled the resonator quality factor, the detector holder was placed inside a shielding pot consisting in a three layers of Cryophy\textregistered\, copper and aluminum (Fig.~\ref{fig:pentola}).

An optical fiber of 600~$\mu$m core diameter was placed on the top of the pot and illuminated the Si substrate from a distance of 13.6~cm. The opposite end of the fiber was connected to a 400~nm LED lamp located at room temperature outside the refrigerator. Given the numerical aperture of the fiber of 0.22 and the distance from the Si surface, photons illuminate the entire 5x5~cm$^2$ surface, thus simulating the crystal scintillation light. A $^{55}$Fe X-ray source was faced to a corner of the Si substrate, on the opposite position with the respect to the KID and the feedline. 

The resonator was excited and probed at the resonant frequency, $f_0=2.333$~GHz. The output signals were fed into a CITLF3 SiGe cryogenic low noise amplifier~\cite{ampli} installed at the 4~K stage of the cryostat, down-converted at room temperature using a superheterodyne electronics and then digitized at a sampling frequency of 500 kSPS~\cite{Bourrion:2011gi}. Time traces up to 12 ms long of the real (I) and imaginary (Q) parts of the transmission S$_{21}$ were acquired with a software trigger. Finally the I and Q waveforms were converted into phase and amplitude variations relative to the center of the resonance loop~\cite{cruciani2018}.
\begin{figure}[t]
\begin{centering}
\includegraphics[width=\linewidth]{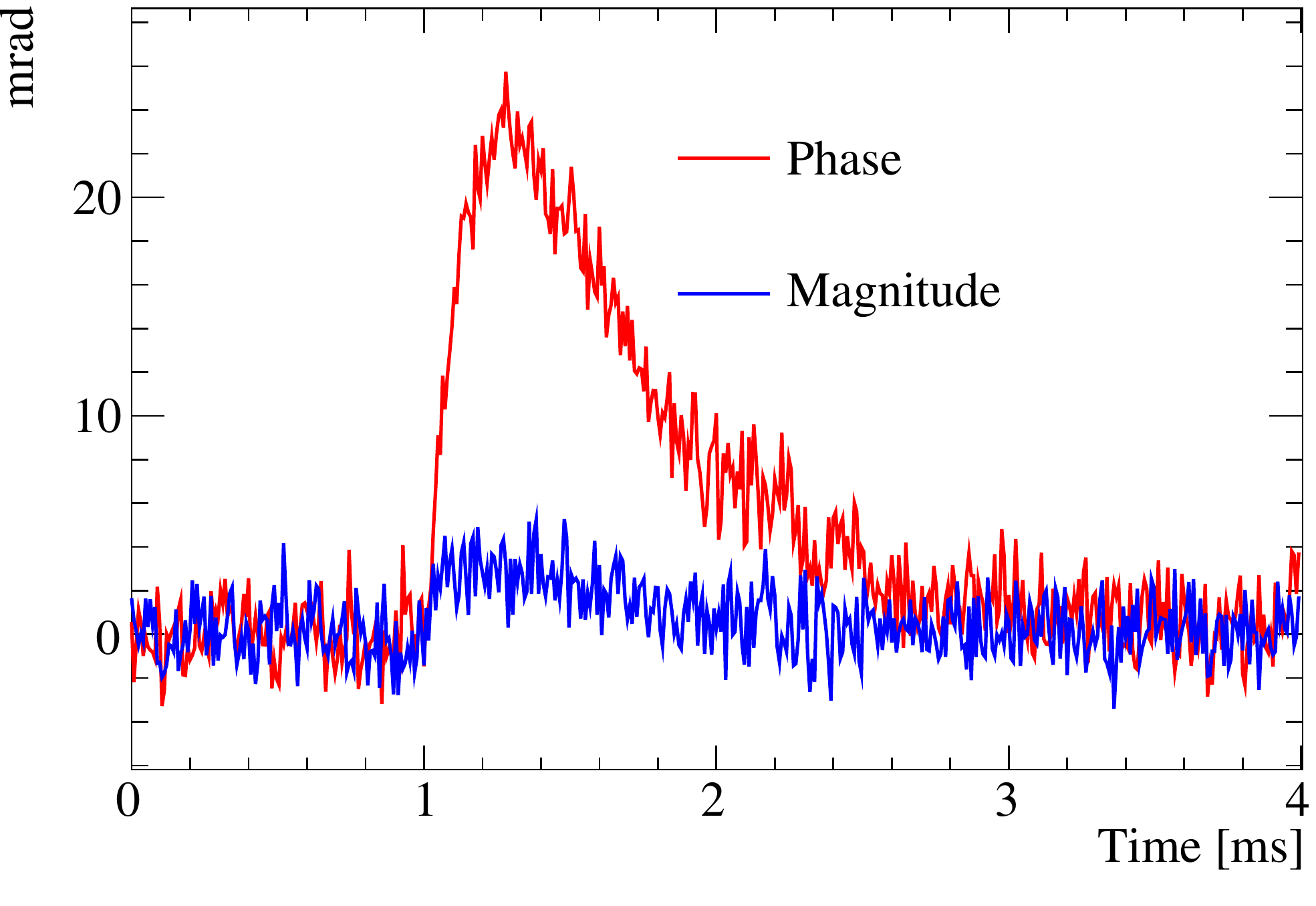}
\caption{Triggered 1.3 keV signal in the phase (red) and amplitude (blue) directions.}
\label{fig:pulso}
\end{centering}
\end{figure}
Figure~\ref{fig:pulso} shows a signal produced by an energy deposition of 1.3~keV in the Si wafer, which is similar to the light signal expected from the \DBD\ of $^{100}$Mo. The average rise time of the signals is 120~$\mu$s, dominated by the phonon propagation in the substrate, while the decay time is 550~$\mu$s, dominated by the recombination of the Cooper pairs~\cite{Martinez:2018ezx}. For the estimation of the signal amplitude, the phase and amplitude waveforms are combined with a bi-dimensional matched filter to maximize the signal to noise ratio~\cite{Bellini:2016lgg}.

\section{Results}
\label{sec:results}
\begin{figure}[b]
\begin{centering}
\includegraphics[width=\linewidth]{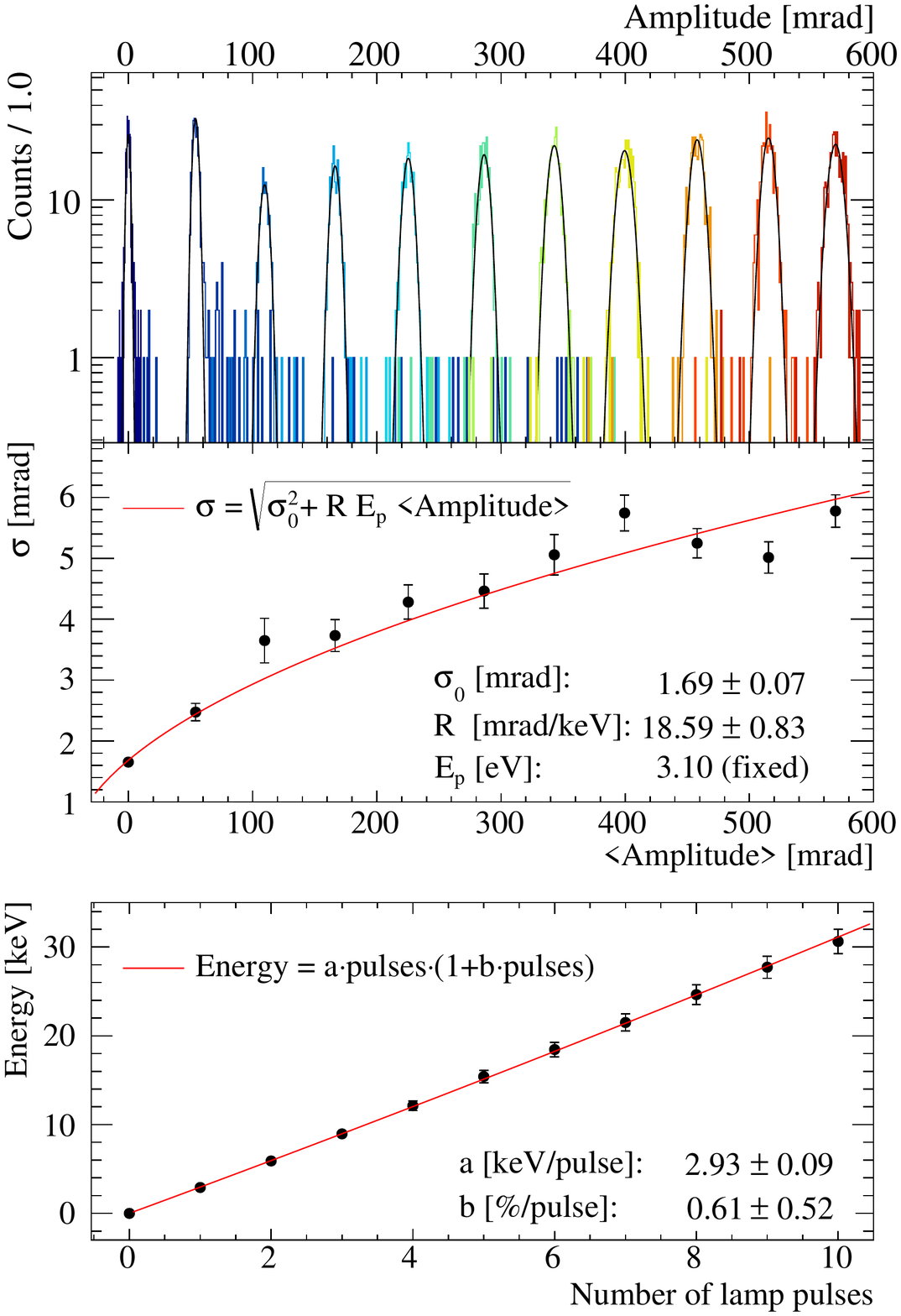}
\caption{Absolute energy calibration using a controlled number of fast LED pulses in order to shine increasing amounts of photons in the substrate. Top: Distribution of the amplitude of the signals after the bi-dimensional matched filter and Gaussian fits to the distributions. Middle: $\sigma$ of the Gaussians as a function of their mean value ($<$Amplitude$>$) and result of the fit to Eq.~\ref{eq:poisson}. Bottom: Calibrated energy as a function of the number of lamp triggers. The fit function shown in the figure is used to estimate the non-linearity $b$ of the calibration procedure and it results negligible.}
\label{fig:abscalib}
\end{centering}
\end{figure}
We used the LED lamp to shine different amount of photons and to perform an absolute energy calibration of the detector~\cite{cruciani2018}. The lamp is controlled with an external trigger that we fire at increasing rate in order to increase the number of photons in a burst. The photon bursts are much shorter than the rise time of the signal, at maximum 2~$\mu$s long, and are therefore seen as a $\delta(t)$ from the KID. Each burst is composed by N  photons of same energy (400~nm, corresponding to 3.1~eV) reaching the Si wafer and absorbed by it. We denote the average number of absorbed photons by $<\rm{N}>$. This process follows the Poisson statistics and therefore the standard deviation is $\sqrt{<\rm{N}>}$. 

As shown in Fig.~\ref{fig:abscalib} (top) for each optical burst, the distribution of the amplitude of the signals is well described by a Gaussian with a standard deviation $\sigma$. 
The predicted trend of $\sigma$ as a function of the mean Amplitude ($<\rm{Amplitude}>$) can be written as the combination of two uncorrelated terms that can be added quadratically, i.e. the poissonian component and the baseline noise energy resolution:
\begin{equation}
\label{eq:poisson}
    \sigma=\sqrt{\sigma_0^2+\rm{R\cdot E_p\cdot <Amplitude>}}
\end{equation}
where $\sigma_0$ is the baseline noise energy resolution, R the energy calibration coefficient and $\rm{E_p}$ the photon energy.
The $\sigma$ vs $<\rm{Amplitude}>$ trend is well described by Eq.~\ref{eq:poisson} as shown in Fig.~\ref{fig:abscalib}-middle and allowed us to extract the calibration coefficient $\rm{R}=18.59\pm0.83$~mrad/keV.
Finally, the bottom panel of the figure shows the calibrated energy as a function of the number of lamp triggers. 

\begin{figure}[t]
\begin{centering}
\includegraphics[width=4.1cm]{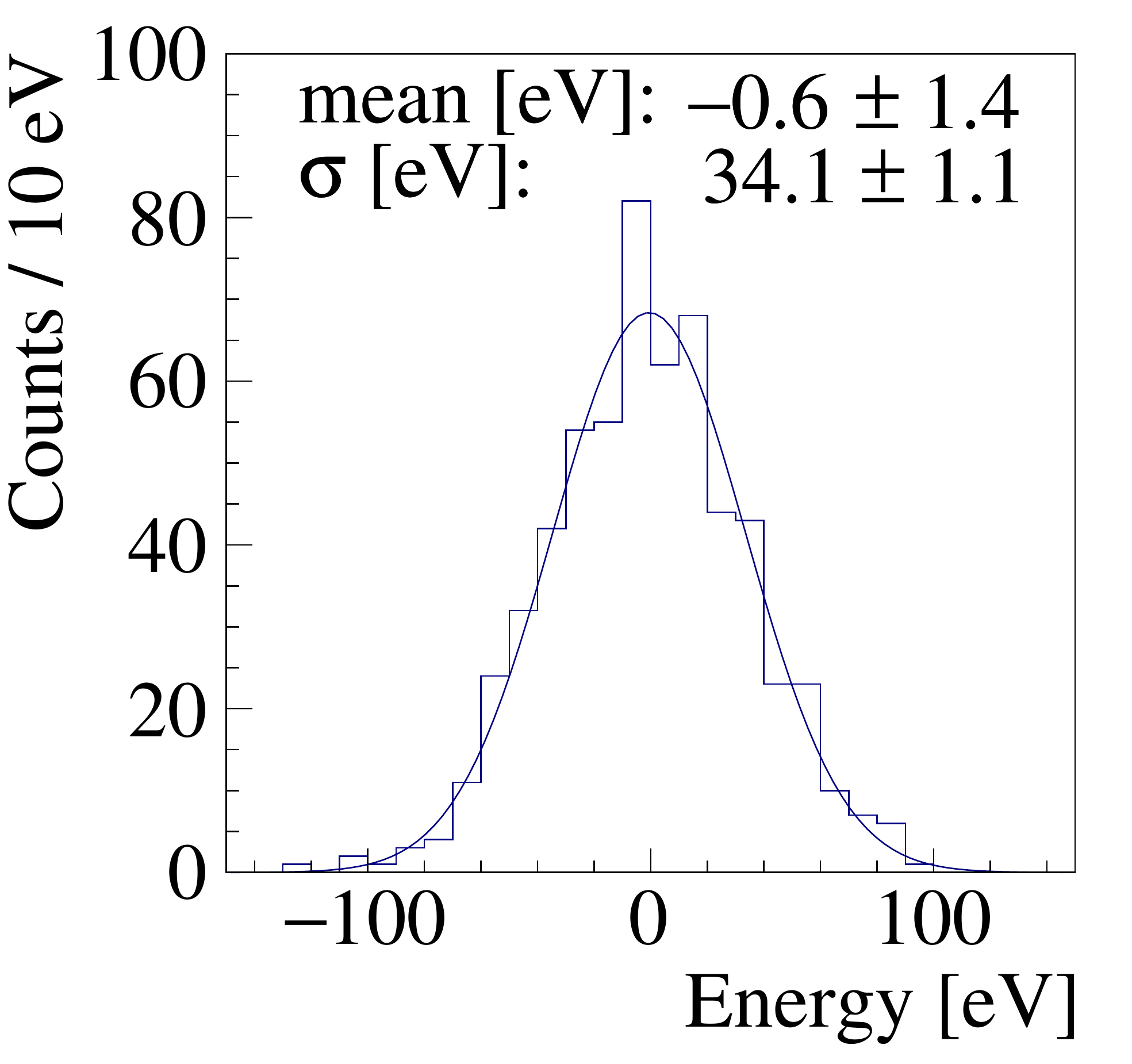}
\includegraphics[width=4.1cm]{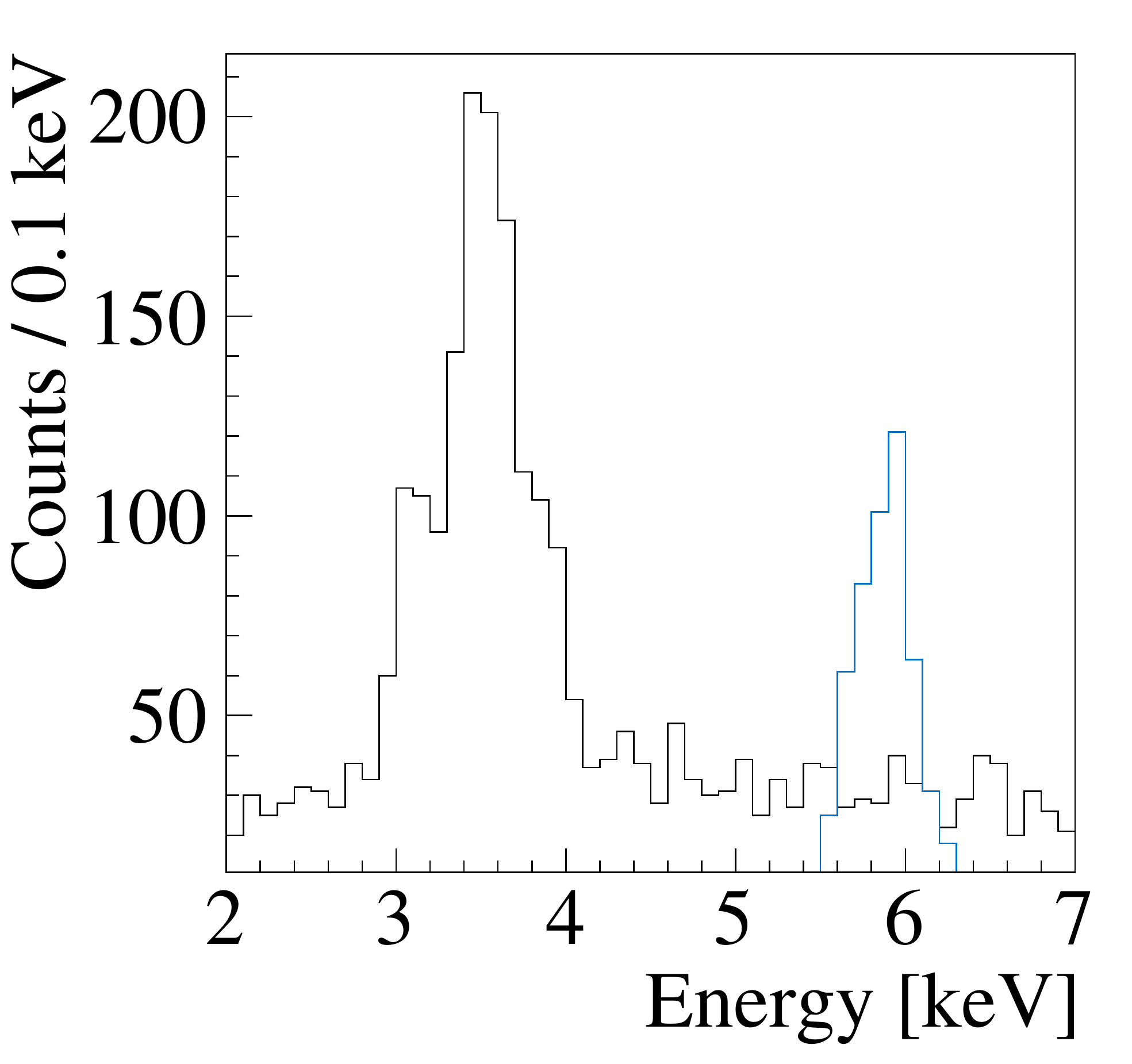}
\caption{Left: Light-calibrated energy distribution of noise triggers with the pulse-tube cryocooler off. Right: light-calibrated energy spectrum acquired with the $^{55}$Fe source (black) and of optical pulses close to the nominal energy of the source (cyan). The peak of $^{55}$Fe is shifted to lower energies because of the position of the source, which fires on a corner of the substrate opposite to the KID position, while the light illuminate the substrate uniformly.}
\label{fig:noise}
\end{centering}
\end{figure}
After calibration the baseline noise energy resolution is 90~eV, a value obtained with the pulse-tube cryocooler on and without any decoupling system to reduce vibrations (see e.g. Ref~\cite{pirro}). These represent the least favorable configuration, since it is well known that the mechanical vibrations induced by dry refrigerators increase the noise  of cryogenic detectors. NTDs and TES detectors, at least in the applications for rare events searches, cannot be even operated without a decoupling system. Indeed, the CUORE infrastructure implments a sophisticated system for the reduction of vibrations~\cite{Nello,cuorecryo} and to emulate such ``ideal'' conditions  we repeated the measurement  with the pulse-tube off and obtained a baseline noise of  $34\pm1(\rm stat)\pm2(\rm syst)$~eV (Fig.~\ref{fig:noise} left), where the systematic error arises from the calibration function. 

Figure~\ref{fig:noise} (right) shows the light-calibrated energy spectrum (black histogram) of the events produced by the $^{55}$Fe X-ray source. The reconstructed energy is significantly smaller the nominal one ($\sim$6~keV), since these devices are sensitive to the position of the energy release~\cite{Martinez:2018ezx} and the source, unlike the fiber that illuminates uniformly the substrate, faces one corner of it far from the KID.
\section{Conclusion}

In this paper we described the technological challenges and solutions in the development of wide-area cryogenic light detectors based on Kinetic Inductance Detectors. We presented the final prototype of the CALDER project, a single KID made of a tri-layer aluminum-titanium-aluminum deposited on a 5$\times$5 cm$^2$ Silicon substrate acting as light absorber. 

This light detector features a rise time of 120 $\mu$s and a vibration-limited noise resolution of 90 eV RMS, matching the requirements of next-generation experiments. We proved that the energy resolution is not limited by the detector itself, rather by the vibrations induced by the pulse-tube cryocooler. In absence of such vibrations (e.g. in fridges such as the CUORE cryostat featuring an active noise cancellation or implementing a decoupling system for the pulse-tube) the energy resolution improve to 34 eV RMS.

Compared to the baseline technology of CUPID, the presented device has a similar energy resolution and an order of magnitude faster response time, a key parameter for background suppression.
Finally, it offers ease in multiplexing and simple fabrication, important aspects for experiments with thousands of channels.

\begin{acknowledgements}
This work was supported by the European Research Council (FP7/2007-2013) under Contract No. CALDER No. 335359 and by the Italian Ministry of Research under the FIRB Contract No. RBFR1269SL. 
The authors thank M. Calvo, M.G. Castellano, C. Cosmelli, A. Monfardini, H. Le Sueur and I.M. Pop for the precious collaboration and the useful discussions during the development of the project.
The authors also thanks the personnel of INFN Sezione di Roma for the technical support, in particular M. Iannone, A. Girardi and S. Casani. 
\end{acknowledgements}

\bibliographystyle{spphys}       

\end{document}